\documentclass[12pt]{iopart}

\usepackage[FIGTOPCAP,small]{subfigure}
\usepackage{graphicx}% Include figure files
\usepackage{dcolumn}% Align table columns on decimal point
\usepackage{bm}% bold math
\usepackage{xcolor}

\newcommand{\bra}[1]{\left< #1 \right\vert}
\newcommand{\ket}[1]{\left\vert #1 \right>}
\newcommand{\pare}[1]{\left( #1 \right)}
\newcommand{\abs}[1]{\left\vert #1 \right\vert}
\newcommand{\cor}[1]{\left[ #1 \right]}

\newcommand{\llav}[1]{\left\lbrace #1 \right\rbrace}

\begin{document}

\title[]{Coherent delocalization: Views of entanglement in different scenarios}
\author{R de J Le\'on-Montiel,$^{1,2}$ A Vall\'es,$^{1}$ H M Moya-Cessa,$^{2}$ and J P Torres$^{1,3}$}
\address{$^1$ICFO - Institut de Ci\`encies Fot\`oniques, Mediterranean
Technology Park, 08860 Castelldefels (Barcelona), Spain}
\address{$^2$Instituto Nacional de Astrof\'{i}sica, \'{O}ptica y Electr\'{o}nica, Calle Luis Enrique Erro 1, Santa Mar\'{i}a Tonantzintla, Puebla CP 72840, Mexico}
\address{$^3$Department of Signal Theory and Communications, Jordi Girona 1-3,
Campus Nord D3, Universitat Polit\`ecnica de Catalunya, 08034
Barcelona, Spain}

\ead{robertoleonm@gmail.com}

\begin{abstract}
The concept of entanglement was originally introduced to explain
correlations existing between two spatially separated systems, that
cannot be described using classical ideas. Interestingly, in
recent years, it has been shown that similar correlations can be
observed when considering different degrees of freedom of a
single system, even a \emph{classical} one. Surprisingly, it has also been
suggested that entanglement might be playing a relevant role in certain biological
processes, such as the functioning of pigment-proteins that
constitute light-harvesting complexes of photosynthetic bacteria.
The aim of this work is to show that the presence of entanglement
in all of these different scenarios should not be unexpected, once
it is realized that the very same mathematical structure can
describe all of them. We show this by considering three different, realistic
cases in which the only condition for entanglement to exist is
that a single excitation is coherently delocalized between the
different subsystems that compose the system of interest.
\end{abstract}

%03.65.Aa   Quantum systems with finite Hilbert space
%03.65.Ud   Entanglement and quantum nonlocality (e.g. EPR paradox, Bell's inequalities, GHZ states, etc.)
%03.67.Mn   Entanglement measures, witnesses, and other characterizations

\maketitle

Entanglement is one of the main traits of quantum theory. For
some, {\em the need to describe even systems that extend over
macroscopic distances in ways that are inconsistent with classical
ideas} \cite{weinberg2013} is a {\em troubling weirdness} of
Quantum Mechanics. Since the publication of the seminal Gedanken
EPR experiment by Einstein, Podolsky and Rosen (EPR)
\cite{EPR1935}, the appearance of the first comments by Bohr about
this subject \cite{bohr1935} and the introduction of the
entanglement concept by Schr\"{o}dinger \cite{schrodinger1935},
innumerable theoretical discussions and experiments related to
this topic have appeared. Arguably the most relevant contribution
to this discussion has been the introduction, fifty years ago now,
of the nowadays well-known Bell inequalities \cite{bell1964}. One
of these Bell-like inequalities, the Clauser-Horne-Shimony-Holt
(CHSH) inequality \cite{CHSH1969}, is surely the most commonly
used in experiments \cite{aspect1982}.

Originally, the application of the concept of entanglement was
restricted to composite systems made up of two spatially separated
subsystems. However, correlations of similar nature to the ones
existing between physically separated subsystems may also exist
when considering different degrees of freedom of a single system \cite{spreeuw1998}.
Indeed, entanglement can be measured in this kind of systems,
provided that one is able to perform independent measurements in
the degrees of freedom involved.

Along these lines, Gadway \emph{et al.} \cite{gadway2009}
demonstrated the presence of entanglement by measuring
correlations in two degrees of freedom (polarization and path) of
a single photon; more recently, Vall\'es \emph{et al.}
\cite{valles2014} enlarged this analysis by considering other
degrees of freedom (orbital angular momentum) and a more general
class of quantum states (mixed states). Violation of  Bell-like
inequalities, a concept related to the presence of entanglement,
has also been used to characterize properties of classical beams
containing many photons, i.e., intense beams. Borges \emph{et al.}
\cite{borges2010} considered coherent beams whose total electric
field writes ${\bf E}({\bf r})=\Psi_H({\bf r}) \hat{\bf e}_H +
\Psi_V({\bf r}) \hat{\bf e}_V $, and used a CHSH inequality to
characterize their coherence properties in one of the two degrees
of freedom involved. Kagalwala \emph{et al.} \cite{kagalwala2013}
added the consideration of partially coherent beams and also
considered the relationship between the degree of Bell inequality
violation and the degree of partial coherence in each degree of
freedom. This so-called {\em non-quantum entanglement} has been
considered as a fundamental tool for investigating important
properties of classical fields
\cite{luis2009,simon2010,eberly2011,ghose2014,toppel2014,aiello2014}.

Here we intend to show that by looking at what is entanglement in
specific, physically realistic scenarios, one can get a better understanding
of what it means to be entangled. We will see that when
considering systems in the single-excitation manifold, entanglement will
always exist as long as the excitation is {\em coherently
delocalized}. We will refer here to coherence as
first-order coherence \cite{glauber1963,glauber1966}, and
delocalization as the fraction of parameter space where the
single-excitation takes place.  In general, coherence and
entanglement do not imply each other, and might address different
aspects of a particular physical system. However, in the
single-excitation manifold \cite{smyth2012}, coherence and
entanglement can become mathematically equivalent. In this
particular regime the presence of entanglement entails coherence
and vice versa, which means that any measure of entanglement is
also a measure of coherence
\cite{sarovar2010,caruso2010,fassioli2010,ishizaki2010}.

The consideration of the single-excitation regime could be seen as
overrestricting our analysis. However, the great majority of
studies of entanglement, both theoretical and experimental, can be
easily demonstrated to belong to this category. Surely, the
implications of being entangled in different contexts might not be
the same, especially when considering subsystems spatially far
away from each other. Notwithstanding, a common conceptual
understanding of entanglement in all of these different scenarios
is still valid and illuminating. For the sake of simplicity, and
keeping a common notation, we will use a quantum language to
describe all scenarios, even when we refer to systems that might
as well be described using classical concepts.

In what follows, we will be more specific about
what {\em single-excitation} regime, {\em
localization} and {\em coherence} mean. With these concepts at
hand, we will explore different scenarios, some of them perfectly
described classically, in which entanglement has been observed.
Even though the appearance of entanglement in some of these cases
might cause certain surprise, we will show that its presence
should not be unexpected if one realizes that such systems can  be
described within the single-excitation manifold, and are therefore
completely analogous to the systems where entanglement is usually
considered.

\section{Entanglement in light-harvesting complexes}
Due to its importance and relevance for explaining and describing
life on earth, photosynthetic light-harvesting complexes have been
a topic of study for decades \cite{blankenship_book}. In recent
years, they have attracted a renewed attention
\cite{ball2011,lambert2013,plenio2013} mainly due to the
experimental observation of long-lived electronic coherences in
the energy transfer process of bacterial and algal
light-harvesting complexes
\cite{engel2007,engel2010,scholes2010,wong2012}. Although the
relevance of some quantum-born concepts, such as entanglement, for
explaining the highly efficient energy transport observed in
photosynthetic systems is still under discussion
\cite{popescu2012,miller2012,roberto2013,mancal2013,roberto_chem},
we will show that the presence of entanglement should not be
unexpected anyway. In the following, we will see that the
appearance of entanglement is a direct consequence
of considering a coherent nature of the
photosynthetic complex, provided the state describing its dynamics
is defined within the single-excitation manifold.

In general, a single excitation in a network of $N$ chromophores
(or sites) can be represented by a density matrix of the form
\begin{equation}\label{density}
\rho = \epsilon \ket{\Psi}\bra{\Psi} + (1-\epsilon) I_{D},
\end{equation}
where
\begin{eqnarray}
\ket{\Psi} &=& \sum_{i}^{N} \alpha_{i}\ket{i}, \label{phot1}\\
I_{D} &=& \sum_{i}^{N} \abs{\alpha_{i}}^{2} \ket{i}\bra{i},
\label{phot2}
\end{eqnarray}
with $\ket{i}$ indicating that the excitation is on site $i$ with
probability $p_i=\abs{\alpha_{i}}^{2}$. The key consideration of
{\em single-excitation} implies that only one site at any time can
be in the excited state. The parameter $\epsilon$ determines the
degree of coherence of the system.

In order to quantify coherence we make use of the degree of
coherence, a function that corresponds to the absolute value of
the normalized first-order coherence function \cite{mandel_book}.
We can then write the degree of coherence as
\begin{equation}\label{coherence}
g_{ij}^{(1)} =
\frac{\mbox{tr}\pare{\rho\sigma_{i}^{\dagger}\sigma_{j}}}{\cor{\mbox{tr}\pare{\rho\sigma_{i}^{\dagger}\sigma_{i}}\mbox{tr}\pare{\rho\sigma_{j}^{\dagger}\sigma_{j}}}^{1/2}},
\end{equation}
where $\sigma_{i}^{\dagger}$ and $\sigma_{i}$ are the raising and
lowering operators for the \emph{i}th site, respectively, and
$\mbox{tr}\pare{\cdots}$ stands for the trace. Making use of Eqs.
(\ref{phot1}) and (\ref{phot2}),  it is straightforward to find
that for the state in Eq. (\ref{density}), the degree of coherence
writes
\begin{equation}\label{visibility}
\abs{g_{ij}^{(1)}} = \epsilon, \; \mbox{for all} \; i \neq j.
\end{equation}
Notice from Eq. (\ref{visibility}) that, depending on the value
$\epsilon$, the degree of coherence can take values from zero,
when there is no coherence, to one, for a fully coherent system.

For the sake of simplicity, and to make more compelling the
comparison with the other cases that will be discussed below, we
restrict our attention to the case of two coupled sites or dimer.
In this scenario, the density matrix in Eq. (\ref{phot2}), in the
basis $\llav{\ket{1},\ket{2}}$, reads
\begin{equation}\label{matrix1}
\rho = \left( \begin{array}{cc}
|\alpha_1|^2 & \epsilon \alpha_1^{*} \alpha_2  \\
\epsilon \alpha_1 \alpha_2^{*}  & |\alpha_2|^2    \end{array}
\right).
\end{equation}
Different measures---such as logarithmic negativity
\cite{caruso2009} and global entanglement
\cite{sarovar2010}---have been used for quantifying entanglement
in light-harvesting complexes. Here, we will quantify the amount
of entanglement present in a two-site system by making use of the
concurrence \cite{hill1997,wootters1998}, which for a density
matrix of the form (\ref{matrix1}) is given by (see Supporting Material of Ref. \cite{yu2009})
\begin{equation}\label{concurrence}
C= 2\;\mbox{max}\llav{0, \epsilon \sqrt{p_1 p_2}} = 2\epsilon
\sqrt{p_1 p_2}.
\end{equation}

Finally, to quantify the degree of excitation's delocalization in
the system given by Eq. (\ref{matrix1}), we introduce a measure of
delocalization that can be defined as
\begin{equation}
\label{localization} D = 2 \sqrt{p_1 p_2}.
\end{equation}
According to Eq. (\ref{localization}), if the excitation spreads
equally over all sites (maximum delocalization), i.e.,
$p_1=p_2=1/2$, one obtains $D=1$; whereas if the excitation
resides in a single site (maximum localization), i.e., $p_1=1$ and
$p_2=0$, or  $p_1=0$ and $p_2=1$, we obtain $D=0$. Notice that
local unitary transformations that affect the excitation in sites
$1$ and $2$ independently do not affect the value of $D$.
Moreover, for a coherent state ($\epsilon=1$), Eq. (\ref{matrix1})
is equivalent to the Schmidt decomposition of the system
\cite{peres_book}, which justifies the validity of $D$ as a good
measure of the excitation's delocalization in the system.

\begin{figure}[t!]
\begin{center}
      \subfigure[]{\includegraphics[width=10cm]{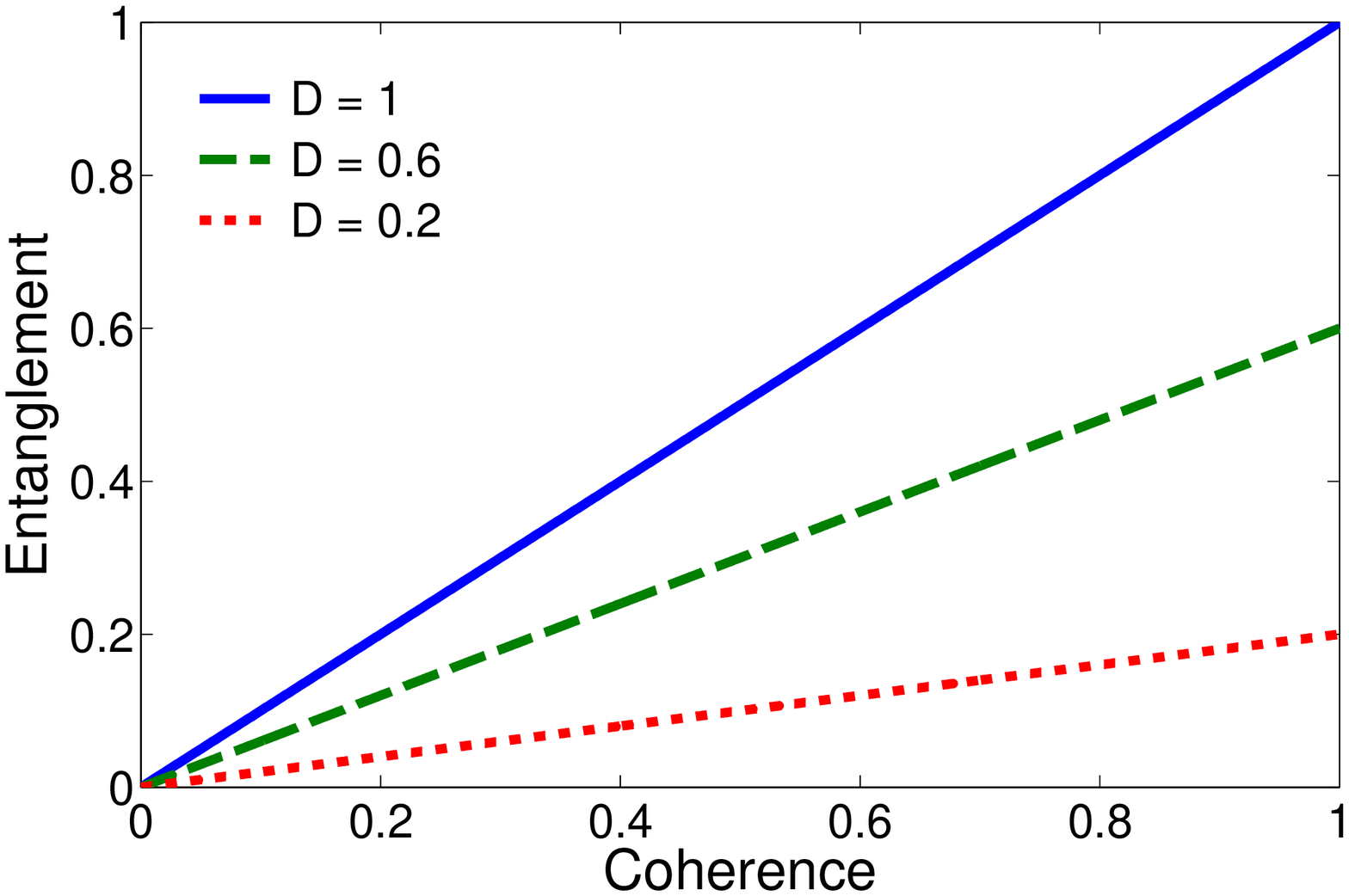}}\\
      \subfigure[]{\includegraphics[width=10cm]{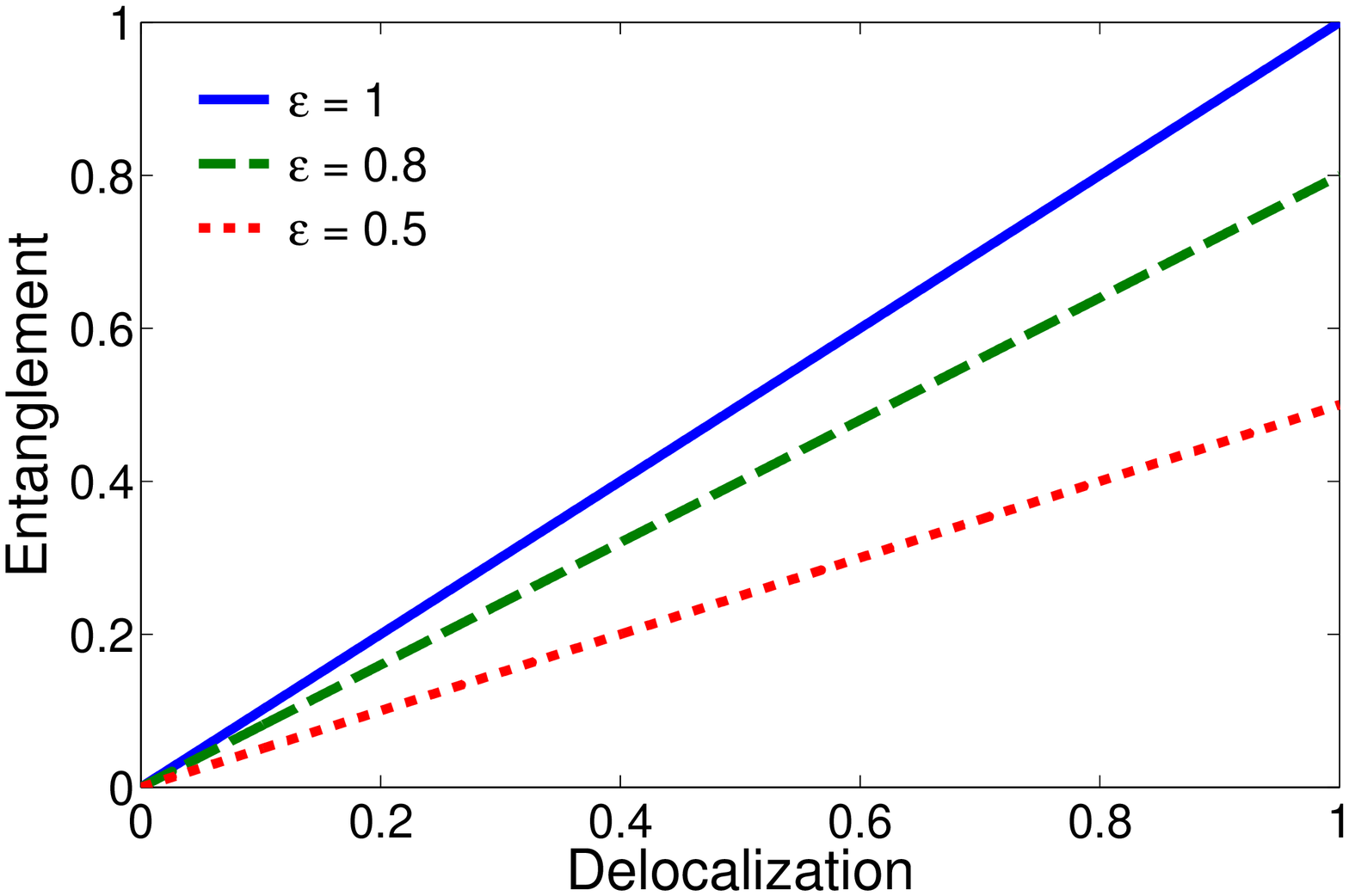}}
\end{center}
\caption{Entanglement, as quantified by the concurrence, as a
function of: (a) degree of coherence $\epsilon$; and (b)
delocalization $D$ of the single excitation. }
\label{fig_entanglement}
\end{figure}

Figure \ref{fig_entanglement}(a) shows the amount of entanglement
(as quantified by the concurrence) as a function of the degree of
coherence $[g_{12}^{(1)}]$ for a fixed value of the degree of
delocalization. For a given delocalization, the degree of
entanglement increases for increasingly larger values of the
coherence. Also, Fig. \ref{fig_entanglement}(b) shows the amount
of entanglement as a function of the degree of delocalization for
a fixed value of the coherence. Notice that also in this case,
increasingly larger values of delocalization provide larger values
of entanglement.

Using Eqs. (\ref{coherence}), (\ref{concurrence}) and
(\ref{localization}), and the results provided in Figs. 1 (a-b),
one can find that
\begin{center}
Entanglement  = Delocalization  $\times$ Coherence.
\end{center}
From this relationship, we can conclude that maximum entanglement
always requires a maximum delocalization of the excitation with
maximum degree of coherence. This situation has been defined by previous authors as
\emph{coherent delocalization} \cite{levi2014}. In contrast, a
maximally delocalized excitation ($D=1$) with no coherence, the
so-called \emph{incoherent delocalization}, produces no
entanglement. Finally, as one can naturally expect, a fully
coherent system with maximum localization ($D=0$) will not exhibit
entanglement.

In the following sections, we will show that similar results can
explain the presence or lack of entanglement in different
scenarios. Even though in the cases that we will describe below
there is not an actual excitation being shared by the subsystems,
we will borrow this term from the present discussion and use it to
describe physical operations that modify certain properties of a
photon, i.e., its polarization or its orbital angular momentum
content. In this way, we will be able to define ``ground'' and
``excited'' states of each subsystem, thus allowing us to
demonstrate the mathematical equivalence of all the cases
considered in this work.

\section{Polarization entanglement in a two-photon state}
The most convenient way to generate entanglement between two
parties is by making use of the nonlinear process of spontaneous
parametric down-conversion (SPDC), where an intense pump beam
interacts with the atoms of a non-centrosymmetric second-order
nonlinear crystal and mediates the generation of paired photons
(signal and idler) that can be entangled in any of the degrees of
freedom that define the parameter space of the photons
\cite{torres2011}.  Polarization entanglement is the most common
type of photonic entanglement, widely used in many quantum
computing and quantum information applications
\cite{bouwmeester_book}, mainly because of the ease with which it
can be generated and manipulated.

In general, the density matrix of the two-photon system can be
written in the same form as Eq. (\ref{density}), with
\cite{torres2011}
\begin{eqnarray}
&\ket{\Psi}& = \alpha_1 \ket{\mbox{H}}_{s}\ket{\mbox{V}}_{i} + \alpha_2 \ket{\mbox{V}}_{s}\ket{\mbox{H}}_{i}, \label{pola1} \\
&I_{D}& = \abs{\alpha_1}^2
\ket{\mbox{H}}_{s}\ket{\mbox{V}}_{i}\bra{\mbox{H}}_{s}\bra{\mbox{V}}_{i} + \abs{\alpha_2}^2
\ket{\mbox{V}}_{s}\ket{\mbox{H}}_{i}\bra{\mbox{V}}_{s}\bra{\mbox{H}}_{i},
\label{pola2}
\end{eqnarray}
where $\ket{\mbox{H}}$ and $\ket{\mbox{V}}$ stand for horizontal
and vertical polarization states, respectively, and $s,i$ are the
commonly used labels for the signal and idler photons. Here, the
values of $\alpha_{1,2}$ depend on specific polarization-dependent
characteristics of the photon-generation process
\cite{torres2011}.

One can easily realize that the two-photon state defined by Eq.
(\ref{pola1}) is equivalent to considering a two-site state in the
{\em single-excitation} manifold by identifying the corresponding
``ground'' and ``excited'' states of each photon (or subsystem).
For this, we can take $\ket{\mbox{V}}_{s,i}$ as the ground states
and $\ket{\mbox{H}}_{s,i}$ as the excited states, so we find that
the state given by Eq. (\ref{pola1}) lives in a Hilbert subspace
where only one of the two photons can be in the
excited state, that is, the single-excitation subspace.

The degree of coherence of this system can be written as
\begin{equation}
g_{H_sV_i;V_sH_i}^{(1)} = \frac{\mbox{tr}\pare{\rho
A_{H_sV_i}^{\dagger}A_{V_sH_i}}}{\cor{\mbox{tr}\pare{\rho
A_{H_sV_i}^{\dagger}A_{H_sV_i}}\mbox{tr}\pare{\rho
A_{V_sH_i}^{\dagger}A_{V_sH_i}}}^{1/2}},
\end{equation}
where $A_{H_sV_i}^{\dagger} =
(a_{H_s}^{\dagger}a_{V_i}^{\dagger})$ and $A_{V_sH_i}^{\dagger} =
(a_{V_s}^{\dagger}a_{H_i}^{\dagger})$, with
$a_{H_s,V_s}^{\dagger}$ and $a_{V_i,H_i}^{\dagger}$ being the
operators that create signal ($s$) and idler ($i$) photons with
horizontal ($H$) and vertical ($V$) polarizations. Using Eqs.
(\ref{density}), (\ref{pola1}) and (\ref{pola2}), one obtains that
the degree of coherence reads
\begin{equation}\label{pola_coh}
\abs{g_{H_sV_i;V_sH_i}^{(1)}} = \epsilon,
\end{equation}
which is a result that one can anticipate from Eq.
(\ref{visibility}). On the other hand, it is easy
to see that $D=2|\alpha_1| |\alpha_2|$.

Concurrence is again used for quantifying entanglement in this
system, as well as Eq. (\ref{localization}) for the excitation's
degree of localization. Notice that, in the present scenario,
maximum delocalization ($D=1$) designates the
case where pairs of photons with polarization
$\ket{\mbox{V}}_{s}\ket{\mbox{H}}_{i}$ are as likely to be
generated as photons with polarization
$\ket{\mbox{H}}_{s}\ket{\mbox{V}}_{i}$. Indeed, the same results
as those discussed in the previous section can be obtained for the
two-photon case, which means that measuring entanglement in this system is fully equivalent to measuring coherence.

Experimentally, the quantum state described by
Eqs. (\ref{pola1}) and (\ref{pola2}) may be generated by using two second-order nonlinear crystals, where degenerate and collinear type-II SPDC can take place. The input pump beam is divided with the help of a beam splitter and illuminates both crystals. The probability of generating two pairs of photons, one pair in each crystal, is assumed to be negligible for sufficiently low values of the pumping power. Then, down-converted photons of each crystal are redirected to a polarizing beam splitter (PBS), where they enter through different input ports. In this way, in each output port of the PBS, horizontally and vertically polarized photons can be detected. The probabilities $p_{1,2}$ that the pair of photons originates in each of the two crystals may be engineered in
several ways. For instance, one can control the phase-matching
conditions, or the amount of pump power, independently in each
crystal, effectively varying $p_1$ and $p_2$, and so $D$. In the
case where all pairs of photons come from a single crystal one
would obtain $D=0$; whereas in the case when the pumping power and
phase-matching conditions are equal in both crystals, one would
have $D=1$. The coherence $\epsilon$ can be controlled by introducing/removing delays between paired photons originating from different crystals, which effectively introduces/erases distinguishability between them \cite{torres2011}.

Finally, it is important to remark that a two-photon entangled state could
also be described by a state of the form $\ket{\Psi} = \alpha_1
\ket{\mbox{V}}_{s}\ket{\mbox{V}}_{i} + \alpha_2
\ket{\mbox{H}}_{s}\ket{\mbox{H}}_{i}$. Note that, in this case,
the signal photon's polarization is rotated, which means that, in
order to remain in the single-excitation manifold, its
corresponding ``excited'' and ``ground'' states should rotate as
well. Using these new states one can obtain the same results as
the ones discussed above. Finally, we highlight that the fact that the density matrix of the
two-photon system lies in the single-excitation subspace allows
one to implement experimental setups, such as the one described in
Ref. \cite{hendrych2012}, in which the degree of entanglement
between the two photons is controlled by directly modifying the
off-diagonal terms of the system's density matrix, that is, the
degree of coherence.

\section{Spin-orbit entanglement in single photons}
The spatial shape of photons, or its orbital angular momentum
(OAM) content, is a degree of freedom that has received increasing
attention in the last few years, because it has opened a new
window, easily accessible experimentally, to explore
high-dimensional quantum spaces encoded in single- or two-photon
systems \cite{gaby2007,franke2008}.

Let us consider the case of a single-photon state in which the OAM
and polarization degrees of freedom are used. It has been shown
that it is possible to generate single-photon states in which the
spatial shape and polarization degrees of freedom are effectively
entangled \cite{nagali2009,valles2014}. In this scenario, the
quantum state of the photon would be described by the so-called
single-photon spin-orbit state, whose
density matrix has the same form as Eq. (\ref{density}), with \cite{chen2010,karimi2010}
\begin{eqnarray}
&\ket{\Psi}& = \alpha_1 \ket{H,-1} + \alpha_2 \ket{V,+1}, \label{spin1} \\
&I_{D}& = \abs{\alpha_1}^2 \ket{H,-1} \bra{H,-1} + \abs{\alpha_2}^2 \ket{V,+1}\bra{V,+1}.
\label{spin2}
\end{eqnarray}
Here, the integer $\pm 1$ corresponds to the value of the OAM index ($m=\pm 1$) of
the photon.

Again, we can see that the single-photon spin-orbit state
lies within the single-excitation manifold by identifying the
``ground'' and ``excited'' states for each subsystem. If we define
$\ket{\mbox{V}}$ and $\ket{-1}$ as the ground states, and
$\ket{\mbox{H}}$ and $\ket{+1}$ as the excited states for the
polarization and OAM degrees of freedom, we can readily find that
Eqs. (\ref{spin1}) and (\ref{spin2}) describe a state that is
equivalent to a Hilbert subspace where only one ``excitation''
in any of the two degrees of freedom can exist, i.e., the
single-excitation subspace.

Following the same procedure as in previous sections, we can
quantify coherence in the single-photon system by writing the
first order correlation function as
\begin{equation}
g_{H-1;V+1}^{(1)} = \frac{\mbox{tr}\pare{\rho a_{H-1}^{\dagger}
a_{V+1}}}{\cor{\mbox{tr}\pare{\rho a_{H-1}^{\dagger}
a_{H-1}}\mbox{tr}\pare{\rho a_{V+1}^{\dagger} a_{V+1}}}^{1/2}},
\end{equation}
where $a_{jm}^{\dagger}$ is the operator that creates a photon
with the polarization state $j=H,V$ and OAM index $m=\pm 1$.

Using Eqs. (\ref{density}), (\ref{spin1}) and (\ref{spin2}) we
thus find that the degree of coherence of the single-photon system
is given by
\begin{equation}\label{spin_coh}
\abs{g_{H-1;V+1}^{(1)}} = \epsilon.
\end{equation}
Finally, for quantifying entanglement in this system, we can make
use of the basis\\
$\{\ket{\mbox{H},+1},\ket{\mbox{H},-1},\ket{\mbox{V},+1},\ket{\mbox{V},-1}\}$ to write
the density matrix of the single-photon system, and find that it
has the exact same form as the one described in Eq.
(\ref{matrix1}). It is then straightforward to obtain that the
concurrence for this state is $C= 2\epsilon \sqrt{p_1
p_2}$.

In experiments, the quantum state described by Eqs. (\ref{spin1})
and (\ref{spin2}) may be generated by making use of a
single-crystal collinear type-II SPDC configuration. In this
configuration, one of the photons is projected into different
polarization states while the remaining photon traverses an
optical device that correlates polarization with OAM
\cite{valles2014}. We can control the values of $p_{1,2}$, and
therefore $D$, by defining a proper polarization-state projection.
For instance, by projecting one photon into the polarization state
$\ket{\mbox{H}}$, the remaining photon would be in the state
$\ket{\mbox{V},-1}$, and therefore $D=0$. Similarly, $D=0$ if we
project the photon into the polarization state $\ket{\mbox{V}}$.
Interestingly, if we project one photon into the state
$\ket{\mbox{H}}+ \ket{\mbox{V}}$, the remaining photon will be in
a quantum superposition of both states, thus giving us a maximum
value of delocalization, $D=1$. Also, as discussed in the previous
case, coherence can be controlled by introducing/removing delays
between the generated pairs of photons.

Finally, from the results discussed above, we can conclude that measuring entanglement in a
single-photon spin-orbit system is the same as measuring
coherence. By identifying that the single-photon spin-orbit state
lies within the single-excitation manifold, we can
anticipate the existence of entanglement between the spin and OAM
degrees of freedom, provided that coherence between them is
preserved. This conclusion has been recently verified by
experiments in which the degree of entanglement between different
degrees of freedom of a single photon is controlled by properly
tuning the degree of coherence ($\epsilon$) \cite{valles2014}.

\section{Conclusions}
Entanglement seems to be a ubiquitous concept that, even though it
was introduced to explain a very specific phenomenon of quantum
theory, it can apply as well to many different scenarios. Here we
have shown that indeed this should not be unexpected, because when
considering correlations between different parties---namely
photons, degrees of freedom or sites---in the important case of
the single-excitation manifold, entanglement is equivalent to
coherence or, more specifically, to coherent delocalization.

We have investigated the conditions for the existence, or lack, of
entanglement in three different systems: a) the process of exciton
transport in photosynthetic light-harvesting complexes, which is
generally modeled as a single excitation propagating in an
$N$-site network, b) the two-photon state generated by means of
spontaneous parametric down-conversion in nonlinear crystals and
c) the coupling between different degrees of freedom of a single
photon. Our results show that even though the
physical scenarios of all the cases considered here are different,
their mathematical equivalence is what allows one to expect and
observe entanglement in each one of them. Furthermore, we have
seen that within the single-excitation Hilbert subspace any
measure of entanglement is equivalent to a measure of the degree
of coherence and localization. This implies that any system that
may be described in a similar manner to the single-excitation
manifold will exhibit entanglement as long as coherence and
delocalization between its subsystems are preserved.

Finally, we have explored the reason why entanglement can even be
observed in classical coherent systems
\cite{gadway2009,borges2010,kagalwala2013}. The analysis presented
here demonstrates that the observation of entanglement, even if
the system can be described classically, should not be unexpected
because the concept of entanglement in the single-excitation
manifold is essentially the same as coherence.

\ack
This work was supported by the program Severo Ochoa, from the
Government of Spain, as well as by Fundacio Privada Cellex
Barcelona. JPT acknowledges support from ICREA, Government of
Catalonia. RJLM acknowledges postdoctoral financial support from
INAOE.

\end{document}